# Reflection of a Broadband Electromagnetic Pulse by Planar Plasma Layer at Normal Incidence

*P.A. Golovinski*[1,2], *V.A. Astapenko*[1], *E.S. Manuylovich*[1]

[1] *Moscow Institute of Physics and Technology, Russia,* [2] *Physics Research Laboratory, Voronezh State University of Architecture and Civil Engineering, Russia*

*e-mail: golovinski@bk.ru*

The problem of electromagnetic pulses propagation in plasma was studied in connection with various problems of radio engineering and radar systems. The description of propagation such signals is based on exactly solvable problems for monochromatic waves. We considered two cases: smooth layer and thin sheet. The conditions for different type of reflection are founded.

## 1. Introduction

The problem of electromagnetic pulses propagation in plasma was studied in connection with various problems of radio engineering and radar systems [1] as well as heating and diagnostics laboratory plasma. In so doing, monochromatic or quasi-monochromatic waves as well as pulses with a characteristic length much more than a carrier wavelength were mainly considered. The use of ultrashort pulses, in which a characteristic length is several wavelengths, for radars makes it possible to obtain an essentially new tool for determination object size and shape [2]. Such signals have super broad spectral band, and in reflection from highly dispersive media, as well as in passage through them, are undergo significant distortions [3, 4]. Super broadband signals are signals, the bandwidth of which exceeds 25% of the carrier frequency value.

In description of propagation such signals, it is already impossible to use freely the geometrical optic methods. Wavelengths in a beam have a very wide spread in values, and the applicability the approximation for high-frequency part of a beam can be changed to a violation its conditions for more long-wave components. In this situation, a large role in understanding the physical nature of the process and in the development methods for its calculation is played by exactly solvable models for a monochromatic field. Within the framework of this approach, it has been possible to solve the problem of reflection an ultrashort pulse at normal incidence on a plasma layer of finite thickness and on plasma with linearly growing density [3]. However, the obtained results do not cover a number of typical important cases that are attributed to the shape of a plasma layer and the electromagnetic pulse form. The present paper is intended to make up for this deficiency with the approach to calculation of a reflected pulse based on exactly solvable problems, which are different from [3].

## 2. Method of calculation

An incident plane wave with arbitrary time dependence at the origin of a plasma layer can be written down in the Fourier representation as

$$E_0(t) = \int_{-\infty}^{\infty} E(\omega) e^{-i\omega t} \frac{d\omega}{2\pi}, \qquad (1)$$

where

$$E(\omega) = \int_{-\infty}^{\infty} E_0(t) e^{i\omega t} dt. \qquad (2)$$

After reflection of a pulse from the layer

$$E_R(t) = \int_{-\infty}^{\infty} A(\omega) E(\omega) e^{-i\omega t} \frac{d\omega}{2\pi}, \qquad (3)$$

where $A(\omega) = a(\omega) e^{-i\varphi}$, $a(\omega)$ and $\varphi(\omega)$ are respectively the modulus of reflection coefficient and the phase shift for a monochromatic wave with a frequency $\omega$. For a transmitted pulse, the following field is quite similar:

$$E_S(t) = \int_{-\infty}^{\infty} B(\omega) E(\omega) e^{-i\omega t} \frac{d\omega}{2\pi}, \qquad (4)$$

where $B(\omega) = b(\omega) e^{-i\phi}$, $b(\omega)$ and $\phi(\omega)$ are respectively the amplitude transmittance and the phase shift in a transmitted wave with respect to an incident wave. Our interest is with extremely short pulses, the representation of which will be considered below.

Let us consider the normal incidence of an electromagnetic wave on plasma. The equation for the monochromatic component of the wave at a frequency $\omega$ in the plane of polarization takes the form [1]:

$$\frac{d^2 E}{dz^2} + k^2(\omega, z) E = 0, \qquad (5)$$

where

$$k^2(\omega, z) = \frac{\omega^2}{c^2} \varepsilon(\omega, z). \qquad (6)$$

Here, $\varepsilon(\omega,z)$ is the dielectric permittivity of a medium, which for plasma in view of the dependence of the electron concentration $N$ on the coordinate $z$ looks like

$$\varepsilon(\omega,z) = 1 - \frac{4\pi e^2 N(z)}{m\omega(\omega - i\nu)}. \qquad (7)$$

The equation of the form (5) coincides with the Schrödinger equation for one-dimensional motion

$$\frac{d^2\psi}{dz^2} + \frac{2m}{\hbar^2}(W - U(z))\psi = 0, \qquad (8)$$

where $W$ is the total energy, $U(z)$ is the potential energy, $\hbar$ is the Planck constant. The known exact solutions of this equation are discussed in detail in quantum mechanics manuals [5-7]. The most demonstrative examples are two extreme cases: a smooth barrier with variable width and height determined by the formula

$$U(z) = \frac{U_0}{\operatorname{ch}^2 \alpha z} \qquad (9)$$

and a very narrow potential barrier

$$U(z) = \beta\, \delta(z) \qquad (10)$$

with a thickness that is much less than characteristic wavelengths in an incident pulse, $\beta$ is the barrier parameter with the dimensionality of squared electric charge.

To go from the results of quantum mechanics to the case of wave propagation in plasma (with neglect of attenuation), we will write the formal equality

$$\frac{\hbar^2 \omega^2}{mc^2}\left(1 - \frac{4\pi e^2 N(z)}{m\omega^2}\right) = 2(W - U(z)), \qquad (11)$$

that is,

$$p = \hbar k = \hbar\omega/c, \qquad (12)$$

$$\frac{\hbar^2 \omega_p^2(z)}{2mc^2} = U(z), \qquad (13)$$

where the momentum of a projectile incident on a barrier is $p = \sqrt{2mW}$, and the plasma frequency varying with the depth in a layer is $\omega_p(z) = \sqrt{4\pi e^2 N(z)/m}$.

For a plasma layer with smooth variation of concentration, we obtain

$$U_0 = \frac{2\pi \hbar^2 e^2 N_0}{m^2 c^2} = \frac{(\hbar \omega_{p0})^2}{2mc^2}. \qquad (14)$$

Here the designation is introduced: $\omega_{p0}(z) = \sqrt{4\pi e^2 N_0 / m}$, and

$$n(z) = \frac{1}{\text{ch}^2 \alpha z}, \qquad (15)$$

where the dimensionless function $n(z)$ defines the electron concentration profile.

For a thin sheet

$$\beta = 2\pi \frac{\hbar^2 e^2 N_2}{m^2 c^2}, \qquad (16)$$

where $N_2 = N_0^{2/3}$ is the two-dimensional concentration of electrons in a thin plasma sheet, when

$$n(z) = \delta(z). \qquad (17)$$

To determine the coefficient of reflection from a layer with smooth variation of electron concentration, we will use the exact solution of a corresponding quantum problem [5] (hereafter we use units, in which $\hbar = m = 1$). The asymptotic form of solution at $z \to -\infty$ is

$$E \sim \exp(-ipz) \frac{\Gamma(ip/\alpha)\Gamma(1-ip/\alpha)}{\Gamma(-s)\Gamma(1+s)} + \exp(ipz) \frac{\Gamma(-ip/\alpha)\Gamma(1-ip/\alpha)}{\Gamma(-ip/\alpha - s)\Gamma(-ip/\alpha + 1 + s)}. \qquad (18)$$

The corresponding reflection coefficient is

$$A = \frac{\Gamma(ip/\alpha)\Gamma(-ip/\alpha - s)\Gamma(-ip/\alpha + 1 + s)}{\Gamma(-s)\Gamma(1+s)\Gamma(-ip/\alpha)}, \qquad (19)$$

where

$$s = \frac{1}{2}\left(-1 + \sqrt{1 - 8U_0/\alpha^2}\right). \qquad (20)$$

For the coefficient of reflection from a thin sheet [7], the expression is obtained:

$$A(\omega) = \frac{\omega_0}{i\omega - \omega_0}, \qquad (21)$$

where $\omega_0 = 2\pi e^2 N_2 / mc$ is the characteristic eigenfrequency of the narrow potential barrier (10). It should be noted that for the solid-state concentration of electrons there is the ratio $\omega_p / \omega_0 \approx 170$ (for silver).

The formulas (19) and (21), together with the inverse Fourier transform, make it possible to obtain the shape of a reflected pulse for thick and thin plasma layers.

### 3. Results of numerical calculations

Let us consider reflections from a planar plasma layer described by the formula (15) of two types of pulses: a corrected Gaussian pulse (CGP) and a cosine wavelet pulse without carrier frequency. The time dependence for the first of them is given by the formula

$$E_{CGP}(t) = \text{Re}\left[-iE_0\left\{\frac{(1+\frac{it}{\omega\tau^2})^2 + \frac{1}{(\omega\tau)^2}}{1+(\omega\tau)^{-2}}\right\}\exp\left(\frac{-t^2}{2\tau^2}\right)\exp(i\omega t + i\varphi)\right], \qquad (22)$$

for a cosine wavelet pulse we have

$$E_{\cos}(t) = \frac{2}{\sqrt{3}\sqrt[4]{\pi}}E_0\left(1-\frac{t^2}{\tau^2}\right)\exp(-t^2/2\tau^2). \qquad (23)$$

The corresponding Fourier transforms of the pulses (22) - (23) look like

$$E_{CGP}(\omega',\omega,\tau,\varphi) = iE_0\tau\sqrt{\frac{\pi}{2}}\frac{\omega'^2\tau^2}{1+\omega^2\tau^2}\left\{e^{-i\varphi-(\omega-\omega')^2\tau^2/2} - e^{i\varphi-(\omega+\omega')^2\tau^2/2}\right\}, \qquad (24)$$

$$E_{\cos}(\omega') = 2\sqrt{\frac{2}{3}}\sqrt[4]{\pi}\,E_0\,\omega'^2\,\tau^3\exp(-\omega'^2\tau^2/2). \qquad (25)$$

In practical calculations, we will assume the pulse amplitude to be equal to one: $E_0 = 1$.

The result of calculation of the change a single-cycle CGP shape in reflection from plasma layers of different thicknesses for $\omega_p = 2\omega$ is shown in Fig. 1.

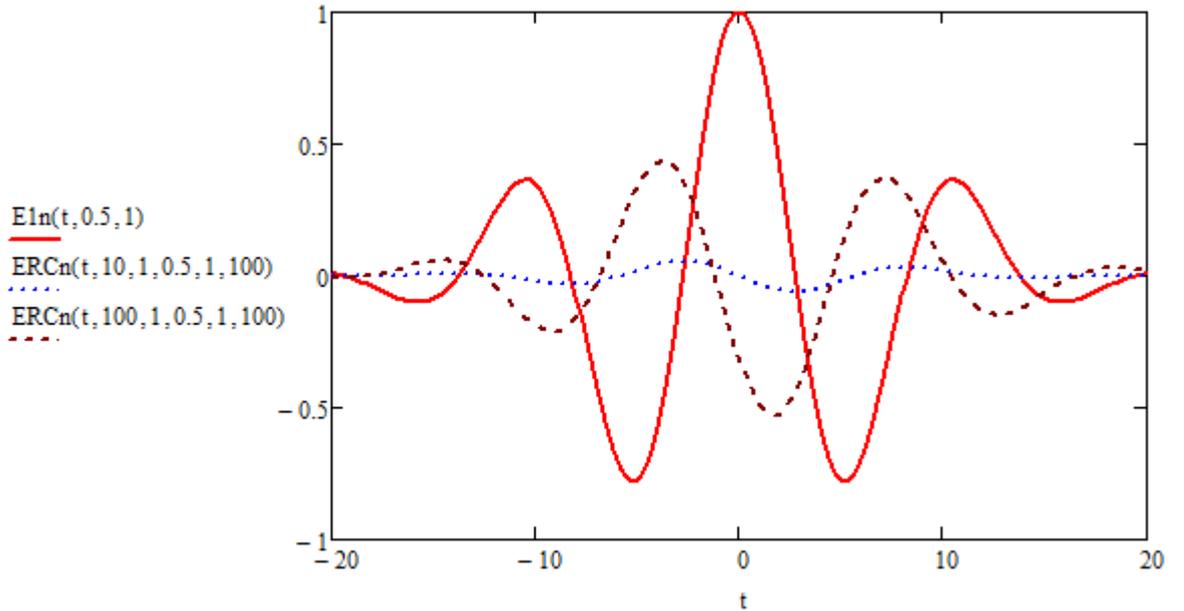

**Fig. 1.** The time dependence of the electric field in a single-cycle CGP reflected from plasma sheets of different thicknesses: solid curve - incident pulse, dotted curve - reflected pulse for $\alpha$ = 0.1 mm$^{-1}$, dashed curve - for $\alpha = 10^{-2}$ mm$^{-1}$; $\omega_p = 10^9$ s$^{-1}$, $\omega_p = 2\omega$

It is seen that with increasing thickness of a layer the amplitude of a reflected pulse increases, that is, the portion of a pulse that passed through the sheet decreases.

The reflection of a single-cycle CGP from plasma layers of different thicknesses for $\omega_p = \omega/2$ is presented in Fig. 2. In this case, since $\omega > \omega_p$, the amplitude of a reflected pulse is considerably less than in the previous situation for the same values of thickness of a plasma sheet. Nevertheless, a reflected pulse exists due to its short duration and accordingly due to the large spectral width, part of which covers the region of partial reflection.

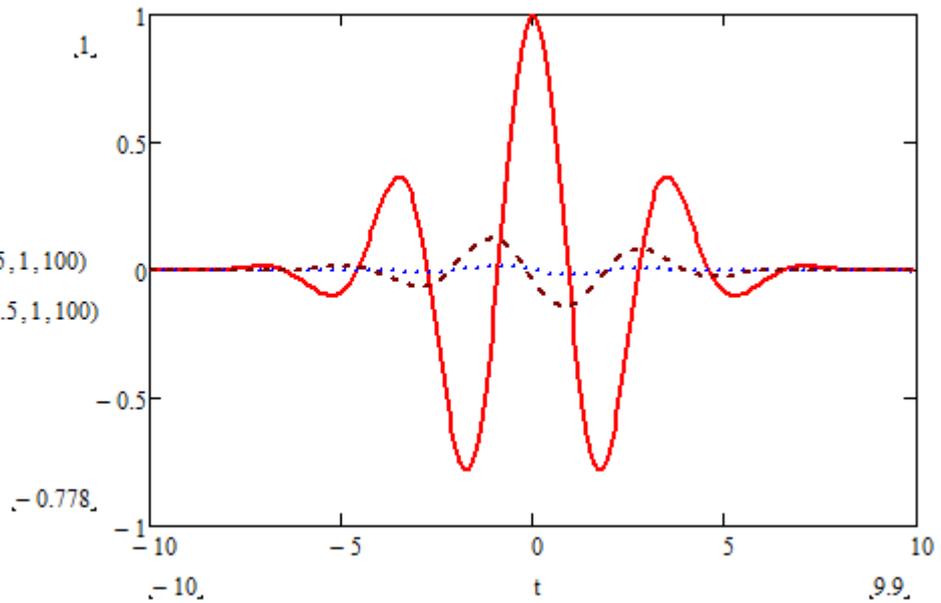

**Fig. 2.** The same as in Fig. 1 for $\omega_p = \omega/2$

Presented in Fig. 3 is the dependence of the reflected CGP shape on the width of a plasma layer for $\omega_p = 2\omega$ and different pulse durations.

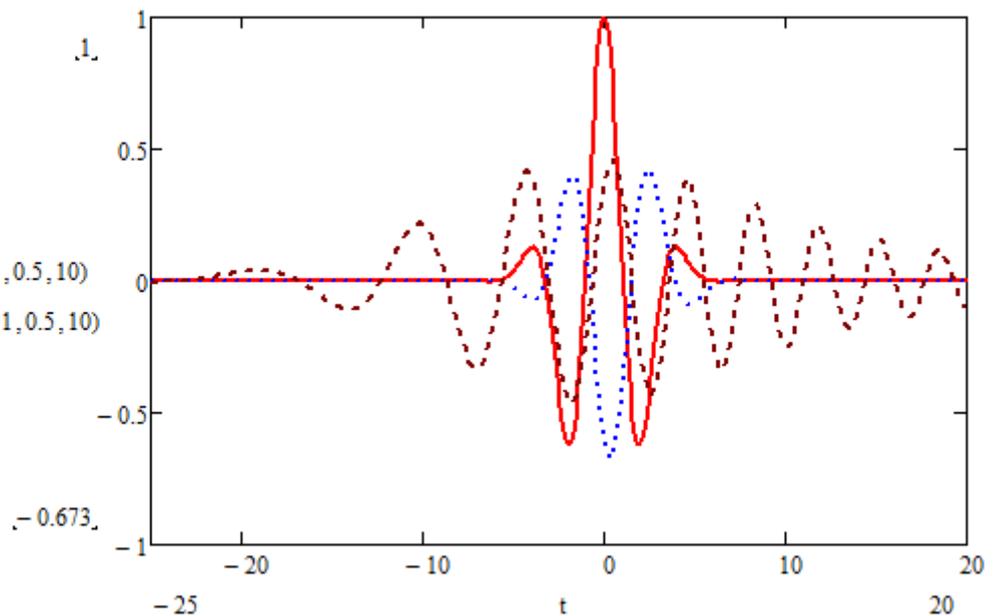

(a) Half-cycle CGP

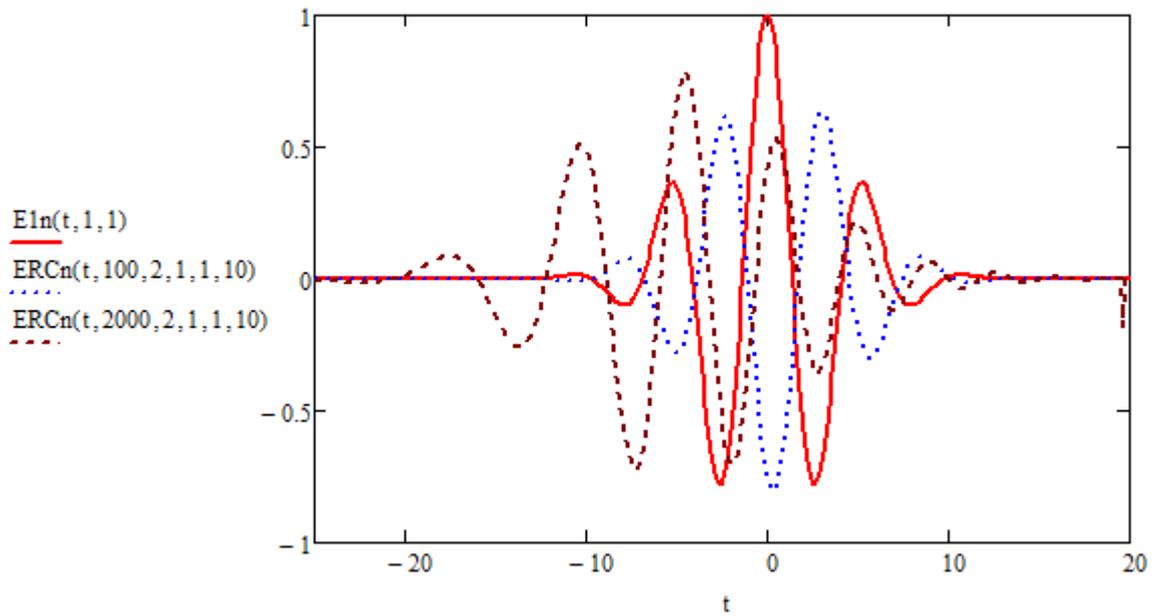

(b) Single-cycle CGP

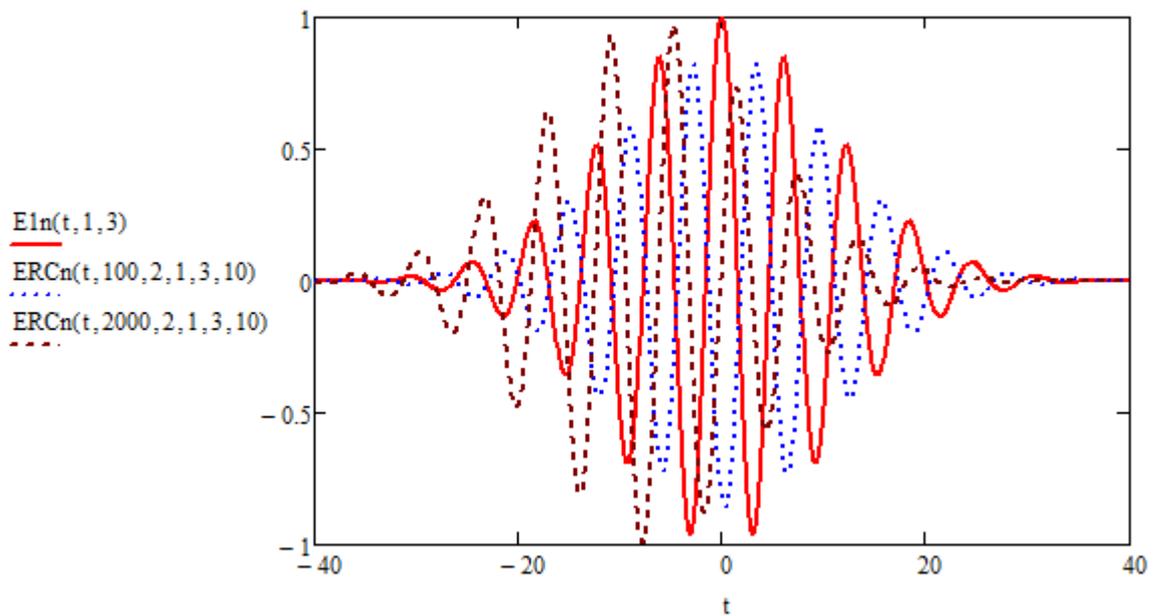

(c) Three-cycle CGP

**Fig. 3.** The evolution of the reflected CGP shape with increasing thickness of a plasma layer: solid curve - incident pulse, dotted curve - $\alpha = 10^{-2}$ mm$^{-1}$, dashed curve - $\alpha = 5 \cdot 10^{-4}$ mm$^{-1}$ for different pulse durations

It is seen that in the case under consideration ($\omega_p > \omega$) chirping, phase shift, and lengthening of a reflected pulse take place with increasing thickness of a plasma layer (with decreasing parameter $\alpha$). In this case with increasing CGP duration these changes become unimportant.

In case of a wavelet pulse (23), the carrier frequency is absent. Its analog is the frequency of the maximum of the pulse spectrum $\omega_{max}$ that is defined by the pulse duration. For a cosine wavelet pulse from the formula (25), we have: $\omega_{max} = \sqrt{2}/\tau$.

The comparison of cosine wavelet pulses of the same duration ($\tau$ = 1 ns) reflected from plasma layers with different plasma frequencies is given by the plots of Fig. 4.

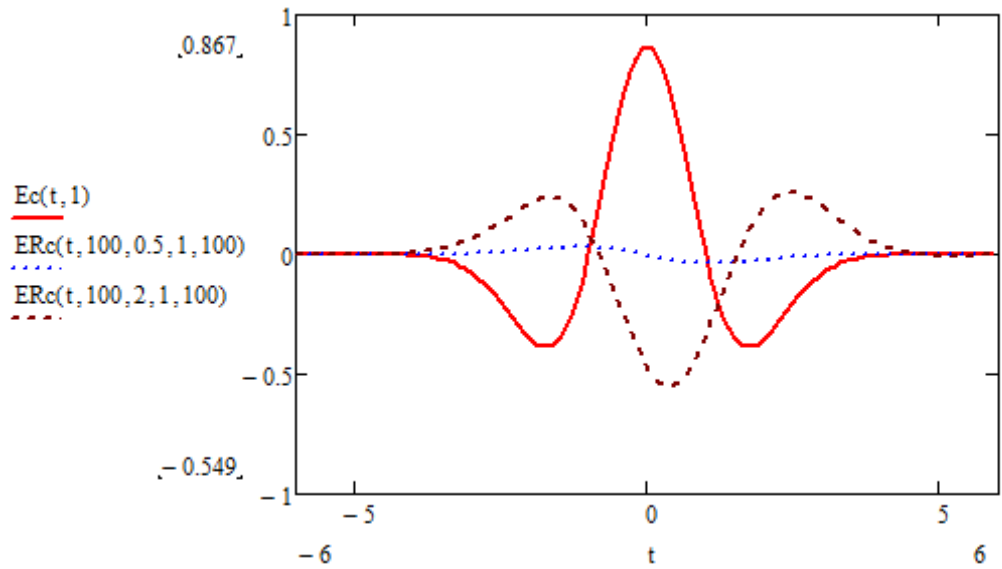

**Fig. 4.** Solid curve - incident cosine wavelet pulse, dotted curve - reflected pulse in case of $\omega_{max} > \omega_p$, dashed curve - reflected pulse in case of $\omega_{max} < \omega_p$, $\tau$ = 1 ns, $\alpha = 10^{-2}$ mm$^{-1}$

As expected from simple physical considerations, with increasing plasma frequency the amplitude of a reflected pulse increases.

The change of the shape of a reflected cosine wavelet pulse with increasing thickness of a plasma layer for the case $\omega_{max} < \omega_p$ is shown in Fig. 5 for different pulse durations. It is seen that, as in CGP reflection (Fig. 3), chirping, phase shift, and broadening of a reflected pulse take place. However, in contrast to a CGP, in case of a wavelet pulse (without carrier frequency) these changes are retained with increasing pulse duration.

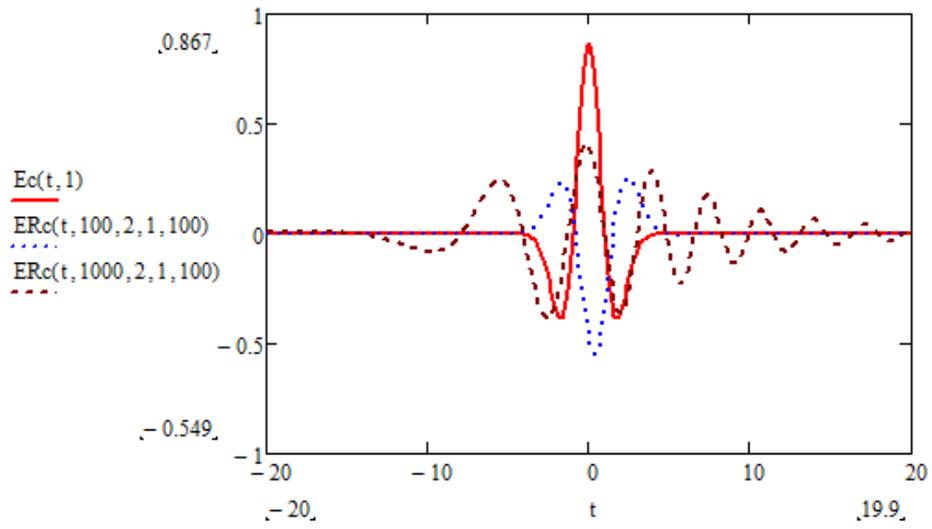

(a) *τ* = 1 ns

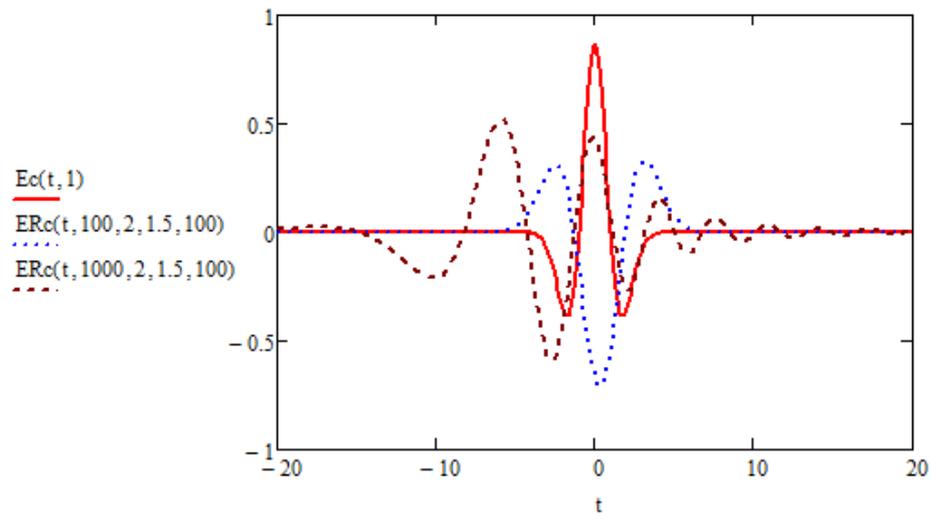

(b) *τ* = 1.5 ns

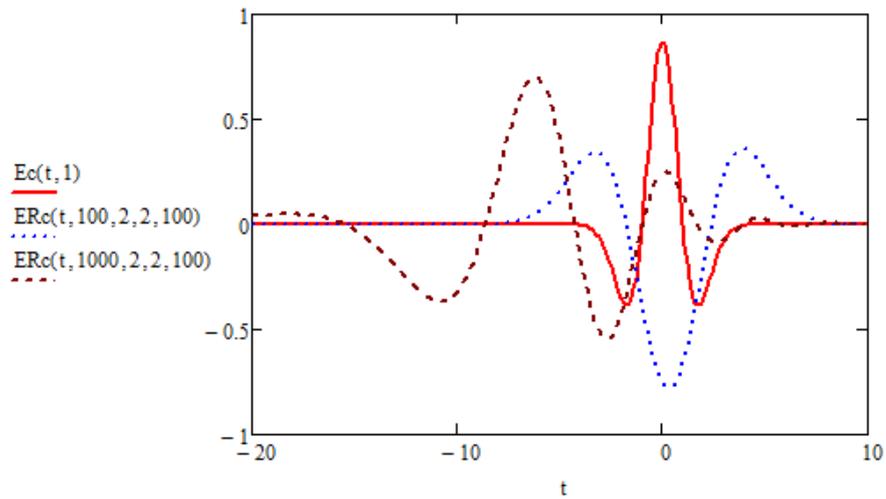

(c) *τ* = 2 ns

**Fig. 5.** The evolution of the reflected cosine wavelet pulse shape with increasing thickness of a plasma layer for $\omega_{max} < \omega_p$: solid curve - incident pulse, dotted curve - $\alpha = 10^{-2}$ mm$^{-1}$, dashed curve - $\alpha = 10^{-3}$ mm$^{-1}$ for different pulse durations

Pulse shapes after CGP reflection from thin plasma sheet with different eigen frequencies are shown in Fig.6 for various pulse durations along with the incident CGP.

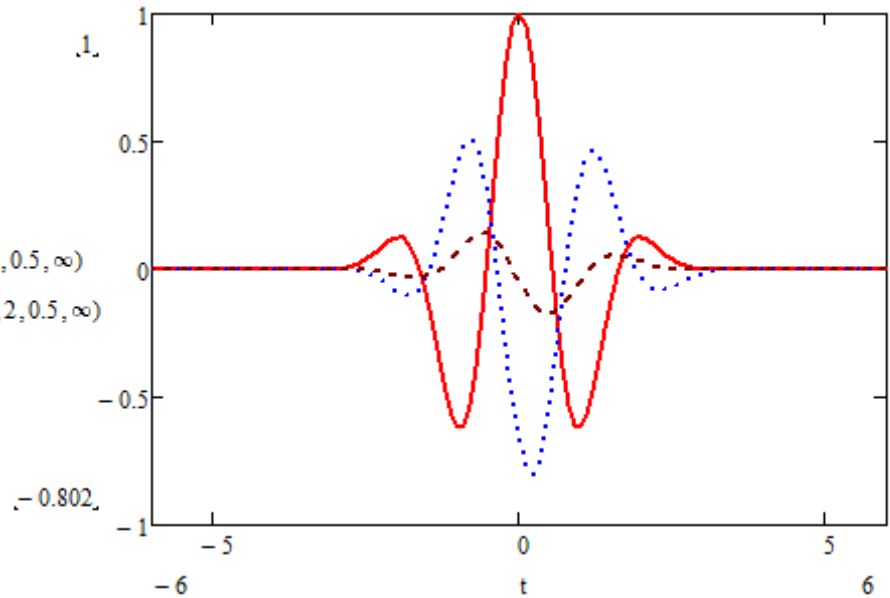

a) Half-cycle pulse

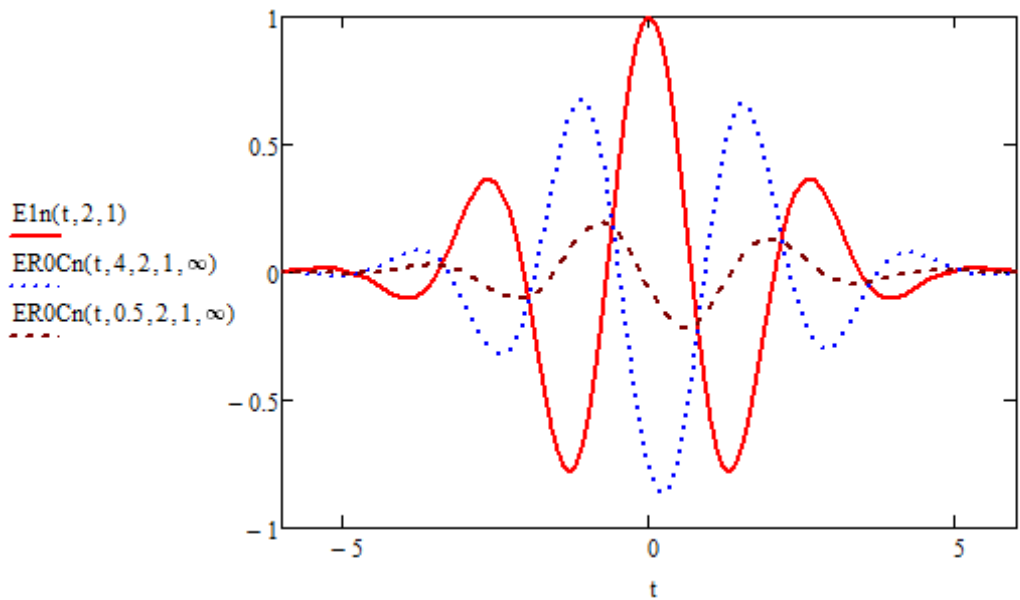

(b) Single-cycle pulse

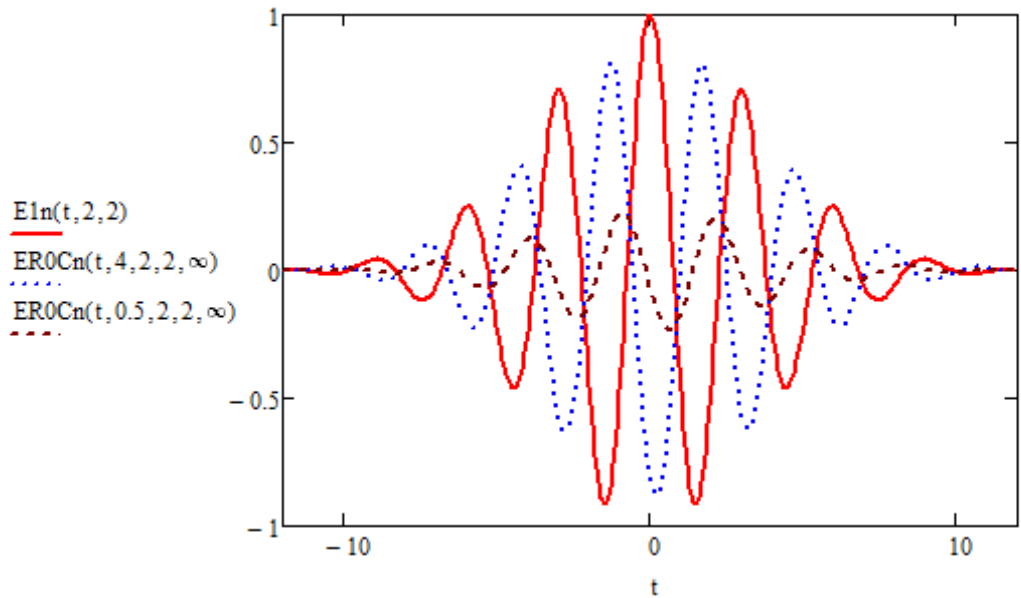

(c) Two-cycle pulse

**Fig. 6.** The change of the shape of a CGP reflected from a thin plasma sheet (the carrier frequency is $\omega = 2$ relative units) for different eigenfrequencies $\omega_0$ of a thin sheet: solid curve - incident pulse, dotted curve - $\omega_0 > \omega$, dashed curve - $\omega_0 < \omega$, and for different pulse durations

As in case of a plasma layer of a nonzero thickness, reflection of a pulse from a thin sheet proceeds more effectively for high eigenfrequencies of such a sheet: $\omega_0 = 2\pi e^2 N_0^{2/3}/mc$. In this case the duration of a reflected pulse practically does not change, and the phase changes in accordance with the relation between the values $\omega_0$ and $\omega$.

**Acknowledgements**

The research was supported by the Russian Foundation for Basic Research (grant No. 015-07-09123 A).